\documentstyle[aps,pra,amssymb,eqsecnum,epsfig,twocolumn]{revtex}

\newcommand{\be}{\begin{equation}}
\newcommand{\ee}{\end{equation}}
\newcommand{\bea}{\begin{eqnarray}}
\newcommand{\eea}{\end{eqnarray}}

\newcommand{\drho}{\delta\rho}

\begin{document}                

\draft
\wideabs{
\title{Zero sound density oscillations in Fermi-Bose mixtures}
\author{P.~Capuzzi and E.~S.~Hern\'andez\cite{byline}}
\address{Departamento de F\'{\i}sica, Facultad de Ciencias
Exactas y Naturales,
Universidad de Buenos Aires, RA-1428 Buenos Aires, Argentina}
\maketitle

\begin{abstract}

Within a mean field plus Random-Phase Approximation formalism, we investigate
the collective excitations of a three component Fermi-Bose mixture of K atoms,
magnetically trapped and subjected to repulsive $s$-wave interactions.  We
analyze both the single-particle excitation and the density oscillation
spectra created by external multipolar fields, for varying fermion
concentrations. The formalism and the numerical output are consistent with the
Generalized Kohn Theorem for the whole multispecies system. The calculations
give rise to fragmented density excitation spectra of the fermion sample and
illustrate the role of the mutual interaction in the observed deviations of
the bosonic spectra  with respect to Stringari's rule. 

\end{abstract}

}
\narrowtext
\section{Introduction}               
\label{sec:intro}

	Shortly after the achievement of Bose-Einstein condensation of dilute,
magnetically confined  alkali atoms, an experimental quest began, seeking to
trap and cool fermionic isotopes of lithium and potassium below their Fermi
degeneracy temperatures $T_F$\cite{potasio1,potasio2,litio1,litio2}.  The
first realization of a weakly interacting, degenerate Fermi gas of K atoms was
reported by DeMarco and Jin\cite{demarco}, and most recently,  a mixture of
bosonic and fermionic Li isotopes has been sympathetically cooled to about
80\% of the corresponding $T_F$\cite{schreck}. On the theoretical side,
various studies of the structure and equilibrium density profiles of
harmonically confined fermions were anticipated in the last
years\cite{teoria}, as well as thermodynamic properties\cite{noro}. The
coexistence of trapped bosonic and fermionic isotopes of a given element
received particular attention
\cite{molmer,amoruso1,vichi1,vichi2,molmer2,viverit,minguzzi1},
 mostly seeking to establish the effects of quantum degeneracy and mutual
interaction on the respective density profiles, energetics and stability
properties.

Since the early days of Bose-Einstein condensates, theoretical and
experimental investigation of the excitation spectra of these objects
constituted a field of active research (see, for example, Refs.
\cite{dalfovo,libro} for substantial reviews) which largely enriched with
reciprocal feedback. In spite of the lack of experimental data concerning
trapped degenerate fermion gases, theoretical study of their collective
oscillations also developed according to two viewpoints. On the one hand, a
hydrodynamic approach derived from the balance equations for the moments
of the Wigner distribution function permitted to compute the sound mode
spectrum of one\cite{amoruso2} and two\cite{amoruso3} hyperfine fermion
species. Such an approach, which can be generalized for anisotropically
trapped systems\cite{csordas} and was applied as well to predict the
characteristics of scissor modes in a superfluid Fermi gas\cite{minguzzi3},
conceals the hypothesis of underlying local equilibrium of the oscillating
gas, i.e., the validity of the Fermi gas equation of state. On the other hand,
since collision rates in a very cold and  strongly dilute gas are small for
moderate numbers of confined particles\cite{amoruso3,bruun2}, one may expect
that a zero sound regime shows up at low excitation energies and temperatures
in these systems; in fact, collective energies of a single-species fermion gas
were estimated within a sum-rule approach\cite{vichi3} for the lowest
multipolarities. Recently, we have proposed a Hartree-Fock (HF) plus
Random-Phase-Approximation (RPA) description of multipolar excitations of two
hyperfine components of $^{40}$K atoms, mutually coupled through a repulsive
$s$-wave interaction\cite{nos}.  The procedure proved adequate to obtain the
trend of collective energies and transition densities for multipolarities $L
= 0, 1, 2$, both for the density and spin density oscillations, and the results
were satisfactorily checked against a simple model, which permits analytical
predictions in the case of identical trapping potentials and particle numbers
for either component.  Moreover, the philosophy turned out to be equivalent to
the one undertaken by Bruun\cite{bruun3}  for Li atoms interacting through an
attractive coupling.

	Due to their coexistence in natural samples, the possibility of
simultaneous trapping and thermal equilibration of both bosonic and fermionic
isotopes of a given alkali within their respective degeneracy regimes, makes
room to  an interesting field for theoretical anticipation, where the study of
excitations of these binary mixtures constitute an open issue. This subject
has not been pursued in depth, as compared with the interest deposited in
structural characteristics and energetics. A three fluid model for trapped
fermions coexisting with both condensed and noncondensed bosons was applied in
Ref.\ \cite{amoruso1} to the study of the temperature dependent profiles, and
in Ref.\ \cite{minguzzi1} to outline a study of density fluctuations within
the hydrodynamic regime. Some features of collisionless, zero sound
oscillations in the Landau limit have been discussed in Ref.\ \cite{viverit}
and in Ref. \ \cite{miyakawa}, a sum-rule approach was applied to extract the
lowest collective zero sound energies of an isotopic mixture of K atoms.  

The purpose of the present work is to contribute to this wide field reporting
calculations of the zero sound spectrum of trapped, weakly interacting K
isotopes - either bosonic $^{39}$K or $^{41}$K, together with two hyperfine
species of $^{40}$K.  The theoretical frame is the mean field-plus-RPA for the
coupled threefold system at zero temperature, which is exposed in
Sec.~\ref{sec:tps} within a simplified scope. The results here presented then
indicate the trend to be expected at very low temperatures, when quantum
degeneracy overcomes the effects of interactions in the fermion sample, and
should be revised if the fermionic gas lies in its superfluid regime. The
procedure can be extended to finite temperatures, at some computational cost.
Specific calculations are analyzed in Sec.~\ref{sec:results}, while the
conclusions and overview of this paper are the subject of
Sec.~\ref{sec:conclu}.

\section{The physical system}
\label{sec:tps}

As anticipated in the Introduction, in this work we study a cold dilute K gas
with three components,  confined in a spherical harmonic potential.  Two of
the components, hereafter called species 1 and 2, are fermions with different
spin projections and the third one is a boson  $B$ species.  Although in
principle, different components subjected to the same magnetic field would
effectively feel  different trapping potentials, we have assumed, for
simplicity, that all atoms experience  the same confinement    
\be
V_{\rm trap} = \frac12 m\,\omega^2\,r^2
\label{pote}
\ee
where $m$ is the atom mass, in other words,  we neglect minor mass differences
among potassium  isotopes, as well as variations in the trap frequency due to
the distinct spin projections. In addition to the external field, the atoms
can interact, but due to the diluteness and temperature of the gas sample it
is enough to consider only $s$-wave interactions\cite{libro}.  Since our gas
includes two types of identical fermions, {\em not} all  $s$-scattering
processes are permitted; in fact, each fermion species is a free Fermi gas
which can only interact with atoms of a different kind.

Taking into account that the $s$-wave interaction can be represented as a
contact potential $V_s = g\,\delta(\mathbf{r-r'})$ with $g=4\,\pi a_s
\hbar^2/m$, being $a_s$ the scattering length of the process, the mean-field
Hamiltonians for the three species can be written as
\bea
H_B &=& {p^2\over 2m} + \frac12 m \omega^2 r^2  + 
g_B\,\rho_B + g_{1}\,\rho_1 + g_{2}\,\rho_2 , \nonumber \\
H_1 &=& {p^2\over 2m} + \frac12 m \omega^2 r^2 + 
g_f\,\rho_2 + g_1\,\rho_B ,\nonumber \\
H_2 &=& {p^2\over 2m} + \frac12 m \omega^2 r^2 
 + g_f\,\rho_1 + g_2\,\rho_B , 
\label{hamilt}
\eea
being $g_B$, $g_1$, $g_2$, and $g_f$ the interaction strength between bosons,
species 1 and B, species 2 and B, and species 1 and 2 respectively. 

These Hamiltonians provide the single particle wave-functions and energies
necessary to find the equilibrium densities $\rho_B,\rho_1,\rho_2$, with
$\rho_B = N_B\,|\phi_0|^2$ and 
\be
\rho_{\sigma} = \sum_{n\,l}(2 l+1)|\psi^{\sigma}_{n\,l}|^2\,
N(\varepsilon_{n\,l}^{\sigma}-
\varepsilon_{F \sigma}),
\ee
for $\sigma$ = 1, 2, being $N(\varepsilon)$ the Fermi occupation numbers at
given temperature $T$, and 
$\varepsilon_{F \sigma}$ the Fermi energy of species $\sigma$, satisfying 
\be
N_{F \sigma} = \sum_{n\,l} (2l+1)\,
N(\varepsilon_{n\,l}^{\sigma}- \varepsilon_{F \sigma}).
\ee

The particle eigenspectrum is the solution of the coupled equations
\begin{mathletters}
\label{excitac}
\bea
H_B \phi_{n\,l\,m} &=& \epsilon_{n\,l} \phi_{n\,l\,m},
\label{gpe}
\\
H_{\sigma} \psi_{n\,l\,m}^{\sigma} &=& 
\varepsilon_{n\,l}^{\sigma}\,\psi_{n\,l\,m}^{\sigma}.
\label{excferm}
\eea
\end{mathletters}

It is worth mentioning that Eq.\ (\ref{gpe}) for $n=0,l=0$ is the
Gross-Pitaevskii equation (GPE) for $N_B$ atoms with chemical potential
$\mu=\epsilon_0$ in an external potential $V=V_{\rm trap} +
(g_1\rho_1+g_2\rho_2)$. The higher modes are the excitation modes of the GPE
which must not be confused with the Hartree-Fock-Bogoliubov (HFB) modes.

\subsection{Collisionless approximation}
\label{sec:phys:coll}

To obtain the spectrum of collective excitations of the compound system we
resort to a random phase approximation, which treats the three species on the
same footing. In a previous work\cite{nos}, we have derived  two  equivalent
sets of two  equations each, for the density fluctuations $\delta \rho^{\sigma
\sigma'}$, being $\sigma, \sigma'$ either 1 or 2. These represent the response
of the density $\rho_{\sigma}$ to an initial perturbation of $\rho_{\sigma'}$.
The diagonal and off-diagonal density fluctuations can be later mapped onto
spin symmetric and spin antisymmetric ones for each species and for the whole
gas. A similar approach in the presence of the bosonic component would lead to
nine equations which include transition densities $\delta \rho^{B B}, \delta
\rho^{\sigma B}, \delta \rho^{B \sigma}$ and corresponding ones for the other
fermion species. In this work, we shall consider identical fermion numbers
$N_{F  \sigma}$  in each species, and we are  mostly interested in total
transitions densities, namely
\bea
\drho_{\sigma}& =& \drho^{\sigma \sigma} 
+ \drho^{\sigma \sigma'}
+ \drho^{\sigma B},
\nonumber  
\\
\drho_{B}& =& \drho^{B B} + \drho^{B \sigma}
+ \drho^{B \sigma'} .
\label{drhoFB}
\eea
Accordingly,  from the full system of equations we can derive a simplified set,
which turns out to be similar to the one developed by Minguzzi and Tosi for
Bose-Einstein condensates at finite temperatures ~\cite{minguzzi4}. The
details are outlined in Appendix A. In this frame, a  density fluctuation
evolves in a perturbed HF potential according to the linearized Hamiltonians
in (\ref{hamilt}).
In terms of the particle-hole (ph) propagators for each spin species in the
HF approximation, $G^{\sigma}_0$, we get for the transition densities
\bea
\drho_{\sigma}({\mathbf r}) &=& \drho_{\sigma}^0({\mathbf r}) 
\nonumber
\\ &&+ 
\int G_0^{\sigma}({\mathbf r},{\mathbf r'})\left[g_{\sigma}\,
\drho_B({\mathbf r'}) +
g_f\,\drho_{\sigma'}({\mathbf r'})\right]d^3r' , 
\nonumber  
\\
\drho_B({\mathbf r}) &=& \drho_B^0({\mathbf r})
\nonumber
\\&&+ 
\int G_0^B({\mathbf r},{\mathbf r'})\left[g_B\,\drho_B({\mathbf r'}) +
\sum_{\sigma} g_{\sigma}\,\drho_{\sigma}({\mathbf r'})\right],
\nonumber \\ \label{drhosigma-drhodef}
\eea
where $\delta\rho_i^0$ is the density fluctuation induced by the external
probe $O^\dag$ in the stationary HF mean-field approximation 
\be
\drho_i^0({\mathbf r}) = \int O^{\dag}({\mathbf r'})\, G_0^i({\mathbf
r},{\mathbf r'})\, d^3r',
\label{drho0}
\ee
and the propagators read
\begin{mathletters}
\label{G0}
\bea
G_0^B&=&N_B \,\sum_{j\neq0}\left\{{\phi_0({\mathbf r'})\,
\phi_0^\ast({\mathbf r})\,
\phi_j({\mathbf r})\,\phi_j^\ast({\mathbf r'}) \over \Omega -(\epsilon_j
-\mu)/\hbar +
i\eta}\right. \nonumber \\
&&-\left.
{\phi_0({\mathbf r})\,\phi_0^\ast({\mathbf r'})\,
\phi_j({\mathbf r'})\,\phi_j^\ast 
({\mathbf r}) \over \Omega + (\epsilon_j -\mu)/\hbar + i \eta } \right\}, 
\label{G0B}
\\
&&\nonumber 
\\
G_0^{\sigma} &=& \sum_{j\,k} 
{N(\varepsilon^{\sigma}_k-\varepsilon_{F \sigma})-
N(\varepsilon^{\sigma}_j-\varepsilon_{F \sigma})
\over \Omega - (\varepsilon_j^{\sigma} - \varepsilon_k^{\sigma})/
\hbar  + i \eta }
\nonumber
\\&& \psi_j^{\sigma *}({\mathbf
r'}) \psi_j^{\sigma}({\mathbf r}) \psi_k^{\sigma *}({\mathbf r}) 
\psi_k^{\sigma}({\mathbf r'}),
\nonumber \\
\label{G0F}
\eea
\end{mathletters}
where each single particle label stands for $n, l, m$, 
and the condensate mode is excluded from  the summation in (\ref{G0B}).

It is clear that in the present situation, i.e., $N_{F1} = N_{F2}$ and
identical trapping field for both hyperfine components, one has a single
fermion spectrum $\varepsilon^{\sigma}_{nl} \equiv \varepsilon^F_{nl}$, one ph
propagator $G_0^{\sigma} \equiv G_0^F$ and same fluctuations $\drho_{\sigma}
\equiv \drho_F$. Due to the spherical symmetry imposed to the trapping
potential and for multipolar operators $O^\dag$, Eqs.\
(\ref{drhosigma-drhodef}) can be rewritten in terms of multipolar components
as  
\bea
\drho^l_{\sigma}(r) =&&\drho^{0\,l}_{\sigma}(r)
+
{4\pi \over 2 l +1}\int r'^2 dr'\,G_{0\,l}^F(r, r')
\nonumber
\\
&&\,[g_{\sigma}\,
\drho^l_B(r') 
 +g_f\,\,\drho^l_{\sigma'}(r')],
\nonumber \\
\drho_B^l(r) = &&\drho_B^{0\,l}(r)
 + {4\pi \over 2 l+1}\int r'^2 dr'\,
G^B_{0\,l} (r,r')
\nonumber
\\
&&[g_B\, \drho_B^l(r')
 + \sum_{\sigma} g_{\sigma}\, \drho_{\sigma}^l(r')],
\label{rpa}
\eea
where the expressions for the $l$-polar propagators $G^i_{0\,l}$ are given 
in Appendix B and the total density fluctuation for type $i$-atoms reads 
\be
\drho_i({\mathbf r}) = \sum_{l} Y_{l\,0}(\hat{r})\,\drho_i^l(r),
\ee
being $Y_{l\,m}$ standard spherical harmonic functions.

 The whole process consists of three stages. First, one has to find the
equilibrium densities, which means to solve the GPE coupled to the fermionic
system of HF equations. This calculation also yields the quasi-particle
spectrum of the fermions. Then, one computes the GPE excitations for the
bosons, and finally, one solves the RPA equations (\ref{rpa}) for the density
fluctuations with given $l$. After computing the transition densities, we
calculate the total susceptibility ($\chi$) and dynamic structure factor ($S$)
of the system as
\bea
\chi^l(\Omega) &=& \chi^l_B + \chi^l_1 + \chi^l_2 ,
\nonumber \\
S^l(\Omega) &=& -\frac{1}{\pi} {\rm Im}\left[\chi^l\right], 
\label{chi-S}
\eea
with $\chi_i^l$ the susceptibility of each component in the interacting system, 
\be
\chi_i^l = \left\{
\displaystyle
\begin{array}{cr}
\displaystyle \int r^{2+l}\drho_i^l(r) dr & l \neq 0 \\
\displaystyle \int r^4 \drho_i^0(r) dr & l = 0   
\end{array}
\right. 
\ee
From the poles of the real part of $\chi^l$ or corresponding peaks in $S$ 
we can extract the excitation energies of the collective modes.

\section{Results} 
\label{sec:results}

In this work, we analyze the spectrum of elementary and collective excitations
for varying concentrations of the fermionic species. For this sake, we
tune the parameters close to the experimental setup of DeMarco {\it et al.}
\cite{demarco}, i.e , we choose $\omega = 2\pi\times 70$ s$^{-1}$,
$a_1=a_2=2.519$ nm, $a_f=8.308$ nm and $a_B=4.292$ nm, corresponding to $^{40}$K
(fermions) atoms and $^{39}$K (bosons).

The computation of the single-particle (sp) spectrum, Eq.\ (\ref{excitac}), was
carried out by expanding their eigensolutions in a basis of harmonic
oscillator functions with 1800 radial wavefunctions. In fact, due to the
spherical symmetry of the trap, we have just diagonalized several small
systems for different  $l$ ($l\lesssim 60$).  Convergence was achieved with an
accuracy of at least $10^{-10}$ in the energies, and the coincidence between
$\mu$ and $\epsilon_0$ was checked to about $10^{-5}$.  Then, the numerical
solution of Eq.\ (\ref{rpa}) for each $\Omega$ and $L$ was performed by
discretizing the problem in a spatial mesh containing up to 250 points and
extending  up to $R \approx 30\, {\mu}$m (approximately fifteen oscillator
sizes); the resulting linear system was solved with standard routines from the
{\tt LAPACK} Fortran library.  

In the next subsections we analyze both the quasi-particle spectrum and the
selfconsistent RPA collective modes.  We use the notation $N_F$ for either
$N_{F\sigma}$ and $g_{\sigma}$ for the common coupling $g_1 = g_2$.

\subsection{Hartree-Fock and GPE excitations}

Within the mean field approach to the coupled system we calculated the
quasi-particle  excitations according to Eqs. (\ref{excitac}).  The main
features of the HF fermionic energies are the following: for low numbers of
fermions in the trap, $N_F \lesssim  10^4$, the excitations energies of the
pure fermion system are basically the harmonic oscillator ones. When we turn
on the coupling to the bosonic component, the repulsion with the condensate
increases the Fermi energy and lowers the slope of the dispersion relation.
The change in slope is more significant, the larger the difference $N_B$ -
$N_F$; indeed, it becomes unimportant when these numbers are comparable or
slightly negative.

\begin{figure}
\centering 
\epsfig{file=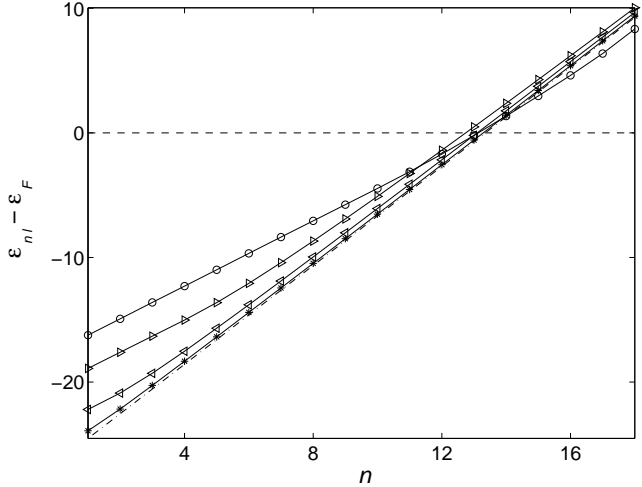,clip=,width=\columnwidth}
\caption{HF fermionic excitation energies (in units of $\hbar\omega$) for 
$N_F= 9880$ and $l=0$.
The different symbols indicate different number of bosonic atoms:  
($\circ$): $N_B=10^6$, $\mu=32.5\hbar\omega$;
($\triangleright$):$N_B=10^5$,$\mu=13.2\hbar\omega$; ($\triangleleft$):
$N_B=10^4$,$\mu=5.7\hbar\omega$; and
($\ast$):$N_B=10^3$, $\mu=2.9\hbar\omega$. The dashed-dotted line shows 
the result for $g_1=g_2=0$. }
\label{fig:hf2}
\end{figure}

As shown by M{\o}lmer\cite{molmer}, for high number of bosons, their kinetic and
interaction energies with the fermions can be neglected in $H_B$, yielding a
Thomas-Fermi approximated (TFA) \cite{stringari96} bosonic particle density 
\be
\rho_B=N_B|\phi_0|^2=\frac{1}{g_B}(\mu-V_{\rm trap})\Theta(\mu-V_{\rm trap}), 
\label{TFA} 
\ee 
where $\Theta(x)$ is the step function.  This  is
translated, after replacement in the fermionic Hamiltonian [cf. Eqs.\
(\ref{hamilt})], into a wider harmonic trap, shifted upwards according to
\bea 
V_{\rm eff} &=& V_{\rm trap} + \frac{g_{\sigma}}{g_B}(\mu-V_{\rm
trap})\Theta(\mu-V_{\rm trap}), 
\nonumber \\ 
&=& 
\left\{
\begin{array}{lc}
\displaystyle \frac{g_{\sigma}}{g_B}\mu + V_{\rm trap}(1-\frac{g_{\sigma}}{g_B}) & {\rm if}\; r<R_B 
\\
& \\ 
\displaystyle V_{\rm trap} & \textrm{otherwise} 
\end{array}\right.
\label{shiftedV} 
\eea
with the Bose radius in the TFA, $R_B = \sqrt{2 \mu/m\omega^2}$. For the
values of the coupling constants here considered, the strength of the
restoring force experienced by fermions inside the bosonic cloud is reduced
by a factor $(1-g_{\sigma}/g_B)\approx 0.41$. Consequently, the  Fermi
radius $R_F = \sqrt{2 \varepsilon_F/m\omega^2}$ acquires an effective value
\be
R_{F\,\rm eff} = \sqrt{2(\varepsilon_F-\mu g_{\sigma}/g_B ) \over
(1-g_{\sigma}/g_B) m \omega^2},
\label{RF}
\ee

which increases with respect to its size in a pure fermion system. The
low-lying quasi-particle excitations ---i.e. those with energies below the
chemical potential $\mu$  of the bosons--- are then expected to be sensitive
to the  renormalized trapping strength, whereas the higher energy ones, which
extend further beyond the central region, are less affected by the effective
trapping. In Fig.\ \ref{fig:hf2} we plot the dispersion relation for
$N_F=9880$, and several values of $N_B$. The corresponding values of the boson
chemical potential are indicated in the caption. We observe that, in agreement
with the effective trapping picture, when the bosonic cloud is large enough,
the low lying modes increase with a smaller slope  than the modes with
energies higher that $\mu$.  Although the fermionic spectrum  is altered by
the interaction, the bosonic counterpart is almost unchanged; indeed, in
addition to the  different number of particles in each isotope, the fermionic
character of the $^{40}$K atoms gives rise to quasi-particle excitations
rather sensitive to variations in the effective trap. The bosonic spectrum is
very different; it actually consists of the GPE single-particle excitations
which will build the collective energies in a RPA calculation.

\subsection{Random-Phase Approximation}

The general aspect of the response (\ref{chi-S}) can be interpreted from its
constituents. Without the boson-fermion coupling, i.e.  $g_1=g_2=0$, the
bosonic response consists of the well-known collective excitations enclosed in
the zero temperature  Gross-Pitaevskii equation (\ref{gpe}), namely, the HFB
spectrum mentioned immediately above Sec. \ref{sec:phys:coll}.  For high
number of bosons, this is well reproduced by Stringari's hydrodynamic
formulation\cite{stringari96}, which provides  an analytic expression for the
multipolar excitation energies.  Although analytic formulae have been also
found  for  systems of interacting fermions ~\cite{minguzzi1,minguzzi3}, they
hold in the hydrodynamic regime and for population ranges far from those we
address here, so we compare our results with earlier numerical calculations
carried within the collisionless regime.  

\begin{figure}
\centering
\epsfig{file=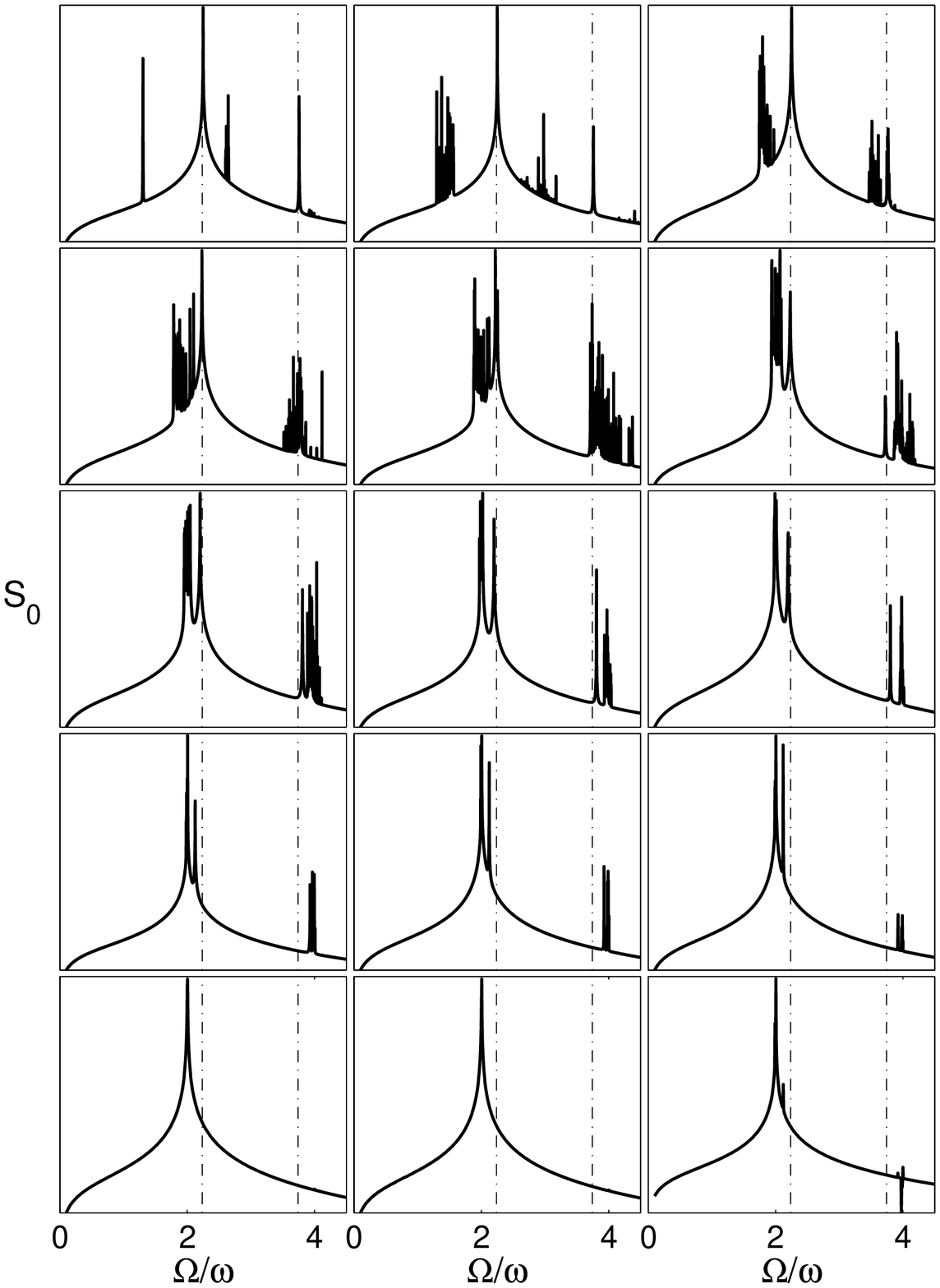,clip=,width=.85\columnwidth}
\caption{Monopolar dynamic structure factor in the RPA approximation (in log scale and arbitrary units). Each row
corresponds to, from top to bottom, $N_B=10^6, 10^5, 10^4, 10^3, 0$; while
each column, from left to right, $N_F= 969, 2925, 9880$. The vertical
dotted lines show  Stringari's energies  \protect\cite{stringari96} for the bosonic system.}
\label{fig:S0_1}
\end{figure}

We have analyzed the dynamic structure factors and fluctuations created by
multipolar external fields with $L=0, 1, 2$.  In Fig.\ \ref{fig:S0_1} we show
the monopolar spectrum for several values of $N_B$  and $N_F$. The essential
features of the monopolar response can be summarized as follows.
\begin{enumerate}
\item{Fixed $N_F$ : for high $N_B$ (higher rows), the spectrum is dominated in
amplitude by the bosonic collective modes with frequencies $\Omega
\approx\sqrt{5},\sqrt{14}$,...[cf. Ref. ~\cite{stringari96}]. However, strong
fragmentation occurs near the fermionic RPA  poles, which can be visualized in
the lowest row. As we lower $N_B$, this fragmentation is reduced and two sets
of two peaks each is distinguished, each set corresponding to the bosonic and
fermionic constituents. The bosonic peak separates from the Stringari spectrum
towards the ideal gas result $\Omega = 2 n \omega$, revealing the unimportant
role of both the boson-boson and boson-fermion interaction in this very dilute
regime.  The fragmentation in the fermionic part cannot be seen due to its 
weakness and the current energy scale.}

\item{Fixed $N_B$: one realizes that increasing $N_F$ not always yields a more
fragmented spectrum. In fact, some peaks appear better defined for larger
values of $N_F$, while other states seem to spread over a larger energy range.
On the other hand, one should keep in mind that the scale in this plot is
governed by the boson strength. If this scale is released, one may observe the
pure fermion RPA peaks (i.e., those with $N_B=0$) within restricted intervals,
with fragmented structures for sufficiently high $N_F$.} 

\end {enumerate}

\begin{figure}
\centering
\epsfig{file=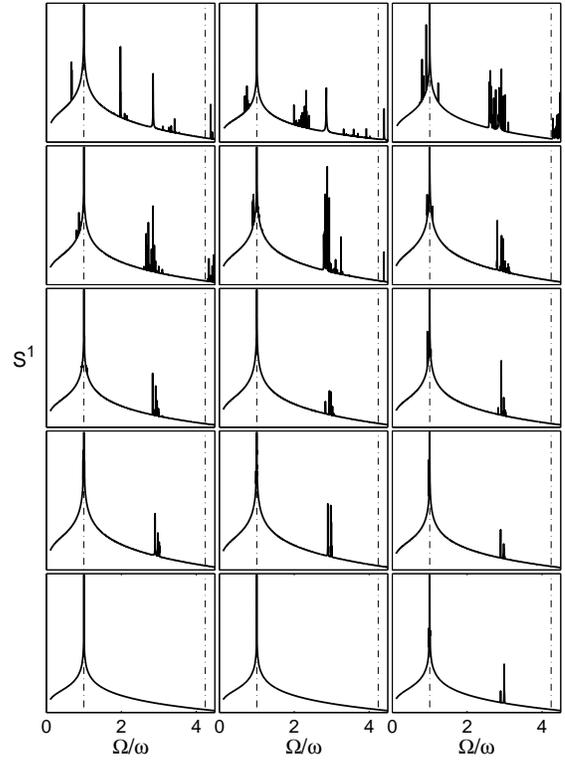,clip=,width=.85\columnwidth}
\caption{Idem Fig.\ \ref{fig:S0_1} for the dipolar response.}
\label{fig:S1_1}
\end{figure}

The results for the dipolar mode ($L=1$) displayed in Fig.\ \ref{fig:S1_1} are
somewhat different. As we can see, for all values of $N_B$ and $N_F$ the
highest peak lies at $\Omega= \omega$, corresponding to a center-of-mass (Kohn
mode) oscillation of the {\em whole} system.  It is known
\cite{nos,kohn,dobson} that both the pure boson and pure fermion systems
possess a well defined Kohn mode within the RPA formulation, which is
recovered in the fully coupled calculation, as can be seen in Fig.\
\ref{fig:S1_1}. In fact, in the numerical calculations the precise location of
this pole has been used to check the computational code.  It should be
mentioned, however, that in a actual experimental setup, the slight differences
between the trap forces for each constituents would distort these modes, giving
rise to coupled oscillation of the components. In addition, the dipolar
spectrum is highly fragmented around  $\Omega=3 \omega$, which coincides with
an oscillator mode for $l=1$.  As in the monopole case, the important
fragmentation seems to disappear as we lower either $N_B$ or $N_F$.     

The quadrupolar mode exhibits the strongest fragmentation. Even for low values
of $N_B$ and $N_F$ the spectrum presents several major peaks whose number
grows at higher densities. At the highest population of all species, the
largest peak corresponds to a boson excitation with $\Omega \approx \sqrt{2}
\omega$ in agreement with the sum-rule estimate of Ref.\ \cite{miyakawa}. As
we diminish the number of bosons the peak is shifted upwards towards the ideal
gas value $\Omega= 2 \omega$.  Another feature not seen in other
multipolarities is the existence of peaks with substantial strength at rather
low energies near  $\Omega \approx 0.3 \omega$.

\begin{figure} 
\centering 
\epsfig{file=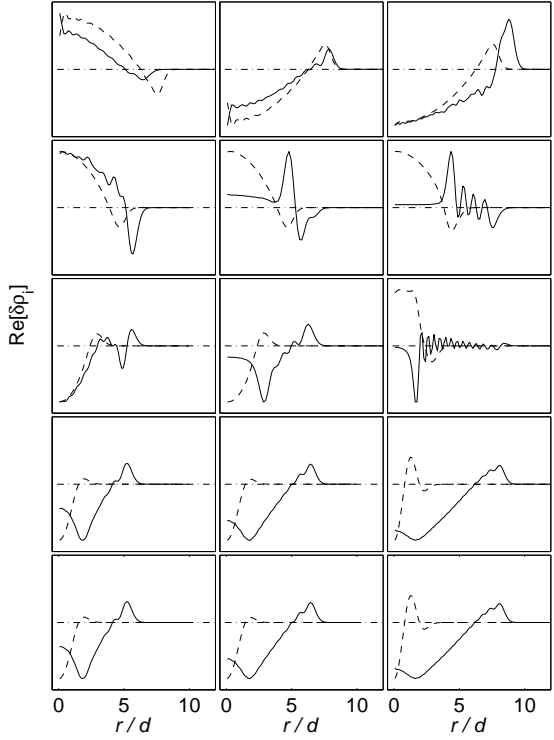,clip=,width=.85\columnwidth}
\caption{Transition densities for the $L=0$ response (in arbitrary units).
The plots show the real parts of the transition densities at the highest peaks
in Fig.\ \ref{fig:S0_1} for each ($N_F$, $N_B$) pair. The solid lines show the
fermionic fluctuation while the dashed ones indicate the bosonic part. Each
plot corresponds to the same ordering as in Fig.\ \ref{fig:S0_1}. 
The positions are expressed in terms of the oscillator size
$d=\sqrt{\hbar/m\omega}\approx 1.9 \mu$m.}
\label{fig:drho0_1} \end{figure}

The shape of the transition densities deserves its own study. In Fig.\
\ref{fig:drho0_1} we depict the monopolar fluctuations corresponding to the
frequencies where the maximun transition probability is attained, each curve
has been independently scaled to fit in the plot. Apart from  differences in
scale, the shapes of the bosonic fluctuations are very similar for all $N_B$
and $N_F$, displaying in any case a volume-like excitation with nonvanishing
density at the origin.  However, two kinds of fermionic excitations can be
seen, a   volume-like one, similar to the bosonic mode, and another which
oscillates  between the boundaries $R_{F\,\rm eff}$ and $R_B$ of the isotopic
particle  densities.  These two shapes can be correlated to the spectrum [see
Fig.\ \ref{fig:S0_1}]: while the volume-like fluctuations occur for
essentially pure boson peaks and in configurations where the bosonic cloud
surrounds the fermionic component, the surface-like mode appears near the pure
fermion excitation. The fermionic atoms extend outwards with respect to the
bosonic density and the excitation bounces between the fermionic and bosonic
density boundaries.  To support this idea we have tabulated the radius of both
densities in Table \ref{tab:param}.

 The monopolar response contains other types of density fluctuations,  some of
them correspond to important poles in the response; one can observe not only
volume-like bosonic excitations, but also surface-like ones, in spatial
counterphase to the fermionic ones for the frequencies shown.   

As expected from the Generalized Kohn Theorem, the dipolar fluctuations at
trap frequency do exist and correspond to a small displacement of the density
yielding a density fluctuation proportional to the gradient of the density,
i.e. 
\be
\drho^{l=1}_i(r,\Omega = \omega) \propto {\partial \rho_i \over \partial r}.
\label{khonmode}
\ee

\begin{figure}
\centering
\epsfig{file=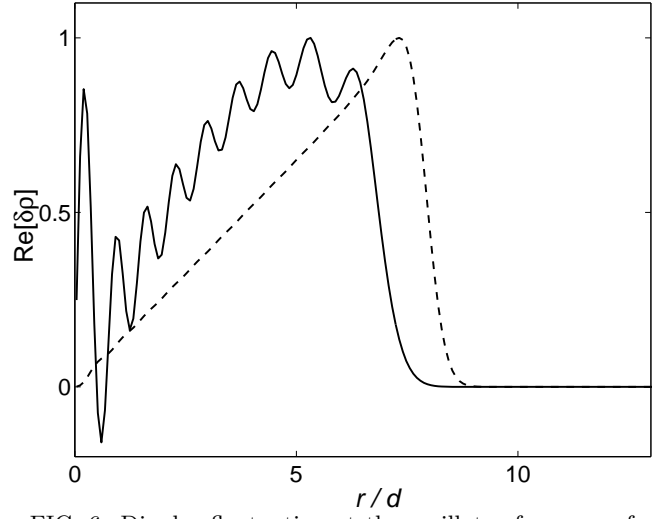,clip=,width=\columnwidth}
\caption{Dipolar fluctuation at the oscillator frequency for $N_B=10^6$ and
$N_F=969$ (in arbitrary units). Solid and dashed lines 
correspond to the fermionic and bosonic fluctuations respectively. Each curve
has been scaled independently.}
\label{fig:drho1_1}
\end{figure}

For high enough $N_B$, according to the TFA, the bosonic density shape inside
the condensate is parabolic, giving a linear $\drho$. This is seen in Fig.\
\ref{fig:drho1_1} where we plot the fluctuation densities corresponding to
these modes for $N_B = 10^6$ and $N_F = 969$. The shape of the fermionic
fluctuation is not as simple, and in fact, should not have a linear dependency
with position since the TFA density profile carries a power $2/3$. Moreover,
the oscillatory shape shown in the plot agrees very well with the
derivative of the fermionic density calculated within the HF approximation. 

Finally, the behavior of the $L=2$ density fluctuations is similar to the
monopolar ones in the sense that for large number of bosons and few fermions,
the highest peak in the response corresponds to a fluctuation which extends
over the bulk of the cloud; whereas for low boson number and/or many fermions,
the fermionic fluctuation is mainly surface-like. The difference with $L=0$
lies in the fact that the quadrupolar oscillation vanishes smoothly at origin.

\section{Conclusions and overview} 

\label{sec:conclu}

In this work we have designed a mean field plus RPA formalism for a trapped
Fermi-Bose mixture, i.e., a three component gas at vanishing temperature.  We
have analyzed both the single-particle excitation spectra and the
collisionless collective modes. On the one hand, the fermionic single-particle
spectra can be  interpreted in terms of an effective confining field induced
by the presence of the bosonic atoms plus the trapping
potential\cite{molmer,amoruso1,vichi1}. This effective field produces a
packing of sp in the lowest branch of the spectrum.   On the other hand, the
bosonic part of the spectrum is mainly unaffected by a moderate fermionic
cloud.  

The poles of the monopolar and quadrupolar response are shifted away from
either the pure fermion or pure boson results and the whole spectrum is highly
fragmented, specially for the quadrupolar probe. In addition, for the dipolar
excitation, the formalism proves to be theoretically and numerically
consistent with the Generalized Kohn Theorem and the density fluctuations are
proportional to the derivative of the corresponding equilibrium densities. The
shape of the monopolar fluctuations reflects  two kind of behaviors. For high
number of bosons, both the fermionic and bosonic fluctuations are volume-like,
extending up to their respective densities edges, while for similar number of
particles, although the bosonic fluctuation remains volume-like, the fermionic
counterpart bounces between  both density borders.   

Finally, this formalism can be extended to finite temperatures where
the system may be found partially degenerate, i.e. with the bosonic 
and/or fermionic components out of the quantum degeneracy regime. This new
regime could give rise to an interesting and rich field of research.

\acknowledgements

This paper was supported by  Grant No.
TW81 from the Universidad of Buenos Aires (UBA). One of us (P.C.) is grateful
to the Universidad de Buenos Aires for financial support.

\appendix
\section{Random Phase Approximation}

	We assume that the trapped, three component  atom system is excited by
the action of an external field, so that ph pairs ($\alpha \beta$)
involving, in principle, both hyperfine fermion species and the bosons, are
created.  We also suppose that a ph effective interaction $V_{\rm ph}^{\alpha
\beta \gamma \delta}$, that permits propagation of elementary
excitations carrying energy $\Omega$ to all orders in the interaction, gives
rise to the dressed RPA propagator for ph pairs  \cite{fetter}
\bea
G^{\alpha \beta}(\Omega) &=& G_0^{\alpha \beta}(\Omega)
\nonumber
\\
& &+ \sum_{\gamma \delta} G_0^{\alpha \beta}(\Omega)\,V_{\rm ph}^{\alpha 
\beta \gamma \delta}\,G^{\gamma \delta}(\Omega).
\label{G}
\eea

We consider longitudinal excitations involving propagation of 
ph pairs of the same spin kind, created by  multipolar operators of the form
\be
O^{\dagger} = \left\{\begin{array}{lc}
\displaystyle r^L\,Y_{L\,0}(\theta,\varphi) & L \neq 0 \\
\displaystyle r^4\,Y_{0\,0} & L = 0, 
\end{array}
\right. 
\ee
and  an effective interaction which couples ph pairs of i) different fermion
species, expressed as $V_{\rm ph}^{\sigma \sigma'} \equiv V_{\rm ph}^{\sigma
\sigma\sigma' \sigma'}$; ii) each fermion species and  the bosons,
$V_{\rm ph}^{\sigma B} \equiv V_{\rm ph}^{\sigma \sigma B B}$ and iii)
bosons,  $V_{\rm ph}^{BB}$.  The free ph propagators involved in longitudinal
density fluctuations are then diagonal in spin space and Eq. (\ref{G}) gives
rise to nine coupled  equations. After convoluting the propagators with the
external field, one gets an equivalent system for the density fluctuations,
which in matrix form can be cast as namely
\bea
\delta \rho^{\sigma \sigma} &=&
\delta \rho_{0}^{\sigma \sigma}
+ G_{0}^{\sigma \sigma}\,(g_f\,
\delta \rho^{\sigma' \sigma}
+ g_{\sigma}\,
\delta \rho^{B \sigma}),
\nonumber 
\\
\delta \rho^{\sigma' \sigma}& = &
G_{0}^{\sigma' \sigma'}\,(g_f\,
\delta \rho^{\sigma \sigma}
+ g_{\sigma}\,
\delta \rho^{B\sigma}),
\nonumber
\\
\delta \rho^{B \sigma}& =& 
G_{0}^{B B}\,(g_{\sigma}\,
\delta \rho^{\sigma \sigma}
+ g_{\sigma'}\,
\delta \rho^{\sigma' \sigma} + g_B\,\delta \rho^{B \sigma}),
\nonumber
\\
\delta \rho^{\sigma B} &=& 
 G_{0}^{\sigma \sigma}\,(g_f\,
\delta \rho^{\sigma' B}
+ g_{\sigma} \,\delta \rho^{B B}),
\label{sistema-c1}
\eea
plus four similar equations corresponding to interchanging $\sigma$ and $\sigma'$, in addition to the modified boson RPA equation
\bea
\delta \rho^{B B} &=&
\delta \rho_{0}^{B B}
 + G_{0}^{B B}\,(g_B\,
\delta \rho^{B B}
 + g_{\sigma}\,
\delta \rho^{\sigma B}
\nonumber 
\\
&& + g_{\sigma'}\,
\delta \rho^{\sigma' B})
\label{sistema-c2}.
\eea

Under the conditions specified in this work, in other words, identical
trapping potentials and same number of fermion of each hyperfine type, the
fermion propagators are identical and one can easily construct the equations
for the total density fluctuations of the fermionic and bosonic atoms, summing
over the hole partner
\be
\drho_{\alpha} = \sum_{\beta} \drho^{\alpha \beta}.
\ee

The final expressions are displayed in Eqs.\ (\ref{drhosigma-drhodef}).

\section{Multipolar decomposition}
To evaluate the free ph propagator we use the 3D basis functions
calculated according to Eq.\ (\ref{excitac}). These eigenfunctions
can be written, for each species, as 
\bea
\phi_{nlm}({\bf r}) &\equiv&R_{n\,l}(r)\,Y_{lm}(\hat{r}),
\nonumber \\
\psi_{nlm}^{\sigma}({\bf r}) &\equiv&R_{n\,l}^{\sigma}(r)\,Y_{lm}(\hat{r}),
\eea
where $Y_{lm}$ are spherical harmonic functions and $R_{n\,l}^{i}$ the
corresponding radial eigenfunctions.

Decomposing the propagators Eqs.\ (\ref{G0}) as
\be
G_0^i({\mathbf r},{\mathbf r'},\Omega) = \sum_{L}G_{0\,L}^i
P_L(\hat{r}\cdot\hat{r'}),
\label{descompG}
\ee
with $P_L(u)$ Legendre polynomials of order $L$.  The
multipolar component $G_{0\,L}^i$ read
\bea
G_{0\,L}^{\sigma} &=& \frac{1}{(4\pi)^2}\sum_{n\,l \atop n'\,l'} 
R_{n\,l}^\sigma(r)\,R_{n'\,l'}^{\sigma}(r)R_{n\,l}^\sigma(r')\,
R_{n'\,l'}^{\sigma}(r')\nonumber \\
&& (2l+1)(2l'+1)|\langle l\,0\,l'\,0 | L 0\rangle|^2\,\chi_{n\,l,
n'\,l'}^{\sigma}(\Omega), 
\eea
being 
\be
\chi_{n\,l, n'\,l'}^{\sigma}(\Omega)= {N(\varepsilon_{n\,l}-\varepsilon_{F\sigma})-
N(\varepsilon_{n'\,l'}-\varepsilon_{F\sigma}) \over \Omega
-(\varepsilon_{n\,l}^{\sigma} -\varepsilon_{n'\,l'}^{\sigma})/\hbar + i\eta},  
\ee
and 
\bea
G_{0\,L}^{B} &=& N_B \phi_0(r)\phi_0(r')\frac{1}{4\pi}{\sum_{n}}'  R_{n\,L}(r)R_{n\,L}(r')
\nonumber \\
&& \left\{{1\over \omega -(\epsilon_{n\,L}-\mu)/\hbar + i\eta}
-{1\over \omega +(\epsilon_{n\,L}-\mu)/\hbar +i\eta} \right\},
\nonumber \\ \label{G0Ldef}
\eea
where the primed summation in Eq.\ (\ref{G0Ldef}) indicate to exclude the
condensate mode, i.e $n \ge 1$ for $L=0$. Similarly, the $l$-polar density
fluctuation read
\be
\drho_i^{0\,L}(r) = {4\pi \over 2 L +1}
\left\{ \begin{array}{cc}
\displaystyle \int r'^{2+L}\,G_{0\,L}^{i}(r') dr' & L\neq 0 \\
\displaystyle \int r'^4\,G_{0\,0}^i(r') dr' & L = 0  .
\end{array}\right.
\ee

\section{Simple proof of the Kohn mode}

The Kohn theorem states that, independently of the interactions, a many-body
system subjected to a harmonic confinement possesses an excited state
corresponding to an oscillation of its center-of-mass at the bare oscillator
frequency $\Omega = \omega$ . Consequently, the fulfillment of Kohn's
theorem is a desired feature of any approximate formulation.

The RPA equations (\ref{sistema-c1}) and
(\ref{sistema-c2}) can be cast in a $9 \times 9$ matrix form, namely
\be
\overline{\overline{\varepsilon}} \cdot \delta \overline{\rho} = \delta
\overline{\rho}_0,
\ee
with $\delta \overline{\rho}= ( \drho^{1 1},\, \drho^{2 1},\, \drho^{B 1},\, 
\drho^{1 2},\, \drho^{ 2 2},\, \drho^{B 2},\, \cdots)^{t}$, and 
\be
\overline{\overline{\varepsilon}}=\left(
\begin{array}{ccc}
{\mathbb A} & 0 & 0 \\
0 & \mathbb{A} & {0} \\
{0} & {0} & \mathbb{A} 
\end{array}
\right),
\ee
being 
\be
\mathbb{A} = \left(
\begin{array}{ccccc}
1 && -g_f\, G_0^1 && -g_1\, G_0^1 \\ &&&& \\
-g_f\,G_0^2 && 1 && -g_2\, G_0^2 \\ &&& & \\
-g_1\,G_0^B && -g_2\,G_0^B && 1 - g_B\,G_0^B
\end{array}
\right) . 
\ee

The collective states of the system are located at the frequencies $\Omega$
which give a singular matrix $\overline{\overline{\varepsilon}}$,
implying a singular  $\mathbb{A}$.  In addition, finding an eigenmode of
$\mathbb{A}$ with zero eigenvalue we prove that $\mathbb{A}$ is singular.     
Moreover, solving this eigensystem is equivalent to solve Eq.\
(\ref{drhosigma-drhodef}) with $\drho_i^0$ set to zero.

We propose the solution 
\be
\delta \overline{\rho}(\omega)=\left({\partial\rho_1\over
\partial{x_j}},\;
{\partial\rho_2\over\partial{x_j} },\;
{\partial\rho_B\over\partial{x_j}}\right)^{t},
\ee
 (for any direction $j$) 
and using the fact that $G_0^i$ is built of
single-particle excitations corresponding to the mean-field Hamiltonians
(\ref{hamilt}) we find \cite{RBGS} 
\bea
\int\! G_0^\sigma({ \mathbf r},{\mathbf r'},\omega)\,\left[g_{\sigma}\,
{\partial\rho_B\over \partial x'_j} + g_f\,{\partial 
\rho_{\sigma'}\over \partial x'_j} \right]d^3r'
&=&
 {\partial \rho_{\sigma}\over \partial x_j}, \nonumber \\
\int\! G_0^{B}({\mathbf r},{\mathbf r'},\omega)\,\left[g_B\,{\partial
\rho_B\over \partial x'_j} + \sum_{\sigma}g_{\sigma} {\partial
\rho_\sigma\over \partial x'_j} \right] d^3r'&=& 
{\partial \rho_B\over \partial x_j},  
\nonumber \\
\eea
which correspond to the requested eigensolution of ${ \mathbb A} \cdot
\delta\bar{\rho}=\mathbf 0$.

\begin{figure}
\centering
\epsfig{file=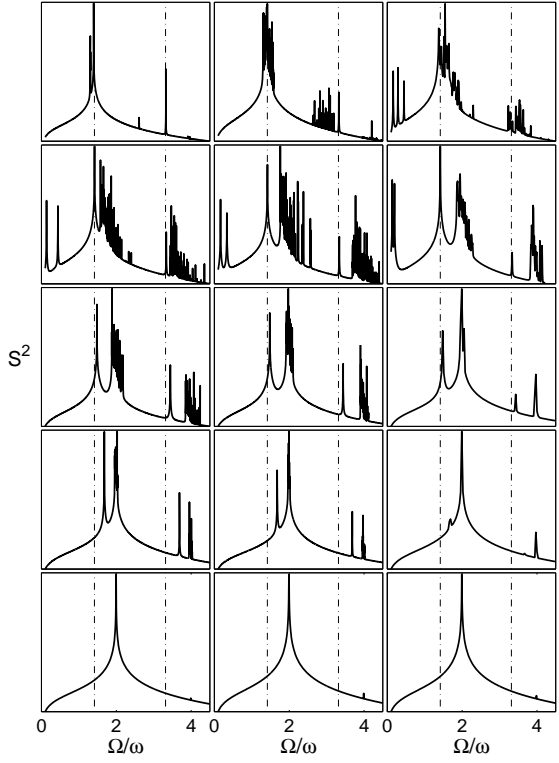,clip=,width=.85\columnwidth}
\caption{Idem Fig.\ \ref{fig:S0_1} for the $L=2$ structure factor.}
\label{fig:S2_1}
\end{figure}

\begin{table}
\caption{Parameters of the stationary densities. $R_B$, $R_{F\,\rm eff}$, 
$\mu$ and $\varepsilon_F$ respectively correspond to the radius of the 
bosonic and fermionic densities and the chemical potential of the bosons 
and Fermi energy.} 
\label{tab:param}
\begin{tabular}{c|dddd}
$N_B$ & $R_B/d$ & $R_{F\, \rm eff}/d$ &$\mu/\hbar\omega$&
$\varepsilon_F/\hbar\omega$ \\
\hline
$N_F=969$ &&&&\\
$10^6$ & 8.3 & 7.0 &32.5& 30.5 \\
$10^5$ & 5.2 & 6.0 &13.0& 20.3 \\
$10^4$ & 3.8 & 6.0 &5.4& 18.5 \\
$10^3$ & 3.0 & 5.8 &2.6& 18.5 \\
\hline
$N_F=2925$ &&&& \\
$10^6$ & 8.4 & 8.4 &32.4&36.1\\
$10^5$ & 5.3 & 7.2 &13.1&27.6\\
$10^4$ & 4.0 & 7.0 &5.5&26.4 \\
$10^3$ & 2.5 & 7.0 &2.7&26.4  \\
\hline
$N_F=9880$ &&&& \\
$10^6$ & 8.0 & 9.2 &32.5&46.0\\
$10^5$ & 5.3 & 8.7 &13.2&40.1\\
$10^4$ & 3.8 & 8.5 &5.7&39.3\\
$10^3$ & 2.4 & 8.5 &2.9&39.3
\end{tabular}
\end{table}

\end{document}